\documentclass[twocolumn,tighten]{aastex61}

\usepackage{graphicx}   
                                                                                          
\newcommand{\co}{\rm CO}
\newcommand{\coa}{CO(1-0)\,}
\newcommand{\coc}{CO(3-2)\,}
\newcommand{\cog}{CO(7-6)\,}
\newcommand{\kps}{\,\textstyle\rm{km~s}^{-1}}
\newcommand{\msun}{\,\textstyle\rm{M}_{\odot}}
\newcommand{\lsun}{\,\textstyle\rm{L}_{\odot}}
\newcommand{\jy}{\,\textstyle\rm{Jy}}
\newcommand{\Kkpspc}{\,\rm{K}\,\rm{km~s}^{-1}\,{\rm pc}^{2}}

\accepted{May 11, 2018}
\submitjournal{ApJ}
\shorttitle{New Molecular Gas Component in SMM02399}
\shortauthors{Frayer et al.}

\begin{document}

\title{The Discovery of a New Massive Molecular Gas Component
  Associated with the Submillimeter Galaxy SMM J02399-0136}

\correspondingauthor{David Frayer}
\email{dfrayer@nrao.edu}



\author{David T. Frayer} \affiliation{Green Bank Observatory, PO Box 2,
  Green Bank, WV, 24944, USA}

\author{Ronald J. Maddalena} \affiliation{Green Bank Observatory, PO
  Box 2, Green Bank, WV, 24944, USA}

\author{R.~J. Ivison}
\affiliation{European Southern Observatory,
  Karl-Schwarzschild-Stra{\ss}e 2, D-85748 Garching, Germany}
\affiliation{Institute for Astronomy, University of Edinburgh, Royal
  Observatory, Blackford Hill, Edinburgh EH9 3HJ, UK}

\author{Ian Smail}\affiliation{Centre for Extragalactic Astronomy,
  Department of Physics, Durham University, South Road, Durham DH1
  3LE, UK}

\author{Andrew W. Blain} \affiliation{University of Leicester, Physics
  \& Astronomy, University Road, Leicester, LE1 7RH, UK} 

\author{Paul Vanden Bout} \affiliation{National Radio Astronomy
  Observatory, 520 Edgemont Road, Charlottesville, VA, 22903, USA}

\begin{abstract}

  We present \coa, \coc, and \cog observations using the Green Bank
  Telescope (GBT) and the Atacama Large Millimeter Array (ALMA) of the
  $z=2.8$ sub-millimeter galaxy SMM\,J02399$-$0136.  This was the
  first submillimeter-selected galaxy discovered and remains an
  archetype of the class, comprising a merger of several massive and
  active components, including a quasar-luminosity AGN and a highly
  obscured, gas-rich starburst spread over a $\sim$25\,kpc extent.
  The GBT \coa line profile is comprised of two distinct velocity
  components separated by about $600\kps$ and suggests the presence of
  a new component of molecular gas that had not been previously
  identified.  The \coc observations with ALMA show that this new
  component, designated W1, is associated with a large extended
  structure stretching 13\,kpc westward from the AGN.  W1 is not
  detected in the ALMA \cog data implying that this gas has much lower
  CO excitation than the central starburst regions which are bright in
  \cog.  The molecular gas mass of W1 is about 30\% of the total
  molecular gas mass in the system, depending on the CO--to--H$_2$
  conversion factor.  W1 is arguably a merger remnant; alternatively,
  it could be a massive molecular outflow associated with the AGN, or
  perhaps inflowing metal-enriched molecular gas fueling the ongoing
  activity.

\end{abstract}




\section{Introduction}

More than 20 years ago, the first submillimeter-selected galaxies
(SMGs) were discovered (Smail et al. 1997; Hughes et al. 1998; Barger
et al. 1998).  The very first SMG detected, SMM\,J02399$-$0136
(SMM02399), was uncovered during the initial surveys of massive
cluster lenses that were used to magnify previously unknown background
SMGs (Smail et al. 1997).  SMM02399 was also the first SMG with a
redshift (Ivison et al. 1998) and the first SMG detected in CO (Frayer
et al. 1998).  The discovery of SMGs revolutionized our understanding
of the high-redshift universe by uncovering a new population of
starburst galaxies that are extremely luminous in the infrared (Blain
et al. 2002; Casey et al. 2014).

Over the last two decades, SMM02399 has been studied in great detail,
and the source has profoundly influenced our understanding of SMGs in
general (Ivison et al. 1998, 2010; Bautz et al. 2000; Vernet \&
Cimatti 2001; Genzel et al. 2003; Lutz et al. 2005; Valiante et
al. 2007; Ferkinhoff et al. 2010, 2011, 2015; Walter et al. 2011;
Thomson et al. 2012; Aguirre et al. 2013).  SMM02399 is compromised of
multiple components, including a bright QSO (L1; Ivison et al. 1998;
Vernet \& Cimatti 2001) which appears to be merging with or is
associated with a nearby extremely red starburst region (L2SW, Ivison
et al. 2010; L1sb, Aguirre et al. 2013).  The optical components of
SMM02399 are shown in Figure~1.

Since SMM02399 is weakly lensed by a foreground cluster (Abell 370 at
z=0.37), differential lensing across the source is not expected to be
significant.  Therefore, one can easily interpret line ratios and the
relative strengths of emission from different regions within SMM02399,
while also benefiting from the boost in brightness and resolution
provided by the foreground lens.  For intrinsic values computed within
this paper, we adopt a magnification factor of $2.38\pm0.08$ (Ivison
et al. 2010).

The nature of the molecular gas in this system is still unclear.
Based on CO(3-2) imaging with the Plateau de Bure Interferometer,
Genzel et al. (2003) argued that the molecular gas lies in a rapidly
rotating disk around the QSO, but more recent CO(1-0) observations
with the Karl G. Jansky Very Large Array (VLA) suggest that the bulk
of the molecular gas is associated with the extremely red starburst
region L2SW (Ivison et al. 2010; Thomson et al. 2012).  Given the
multiple optical components and that the system is associated with a
large diffuse Ly$\alpha$ halo spread over at least 13$\arcsec$ (Vernet
\& Cimatti 2001), it is possible that significant molecular emission
may have been missed in the previous interferometric studies.  In
order to test this idea, we have carried out \coa observations with
the Robert C. Byrd Green Bank Telescope (GBT).  We also present
results based on Atacama Large Millimeter Array (ALMA) \coc and \cog
observations that are helpful in the interpretation of the GBT \coa
data.  Additional results from the ALMA data for SMM02399 will be
discussed in a future paper.

A cosmology of ${\rm H}_0 = 70\kps\,{\rm Mpc}^{-1}$, $\Omega_{\rm
  M}=0.3$, and $\Omega_{\Lambda}=0.7$ is assumed throughout this
paper.  At the redshift of $z=2.808$ for SMM02399, this corresponds to
an image plane angular scale of 7.85\,kpc\,arcsec$^{-1}$ and a
luminosity distance of 23470\,Mpc.

\begin{figure}
\includegraphics[width=0.49\textwidth]{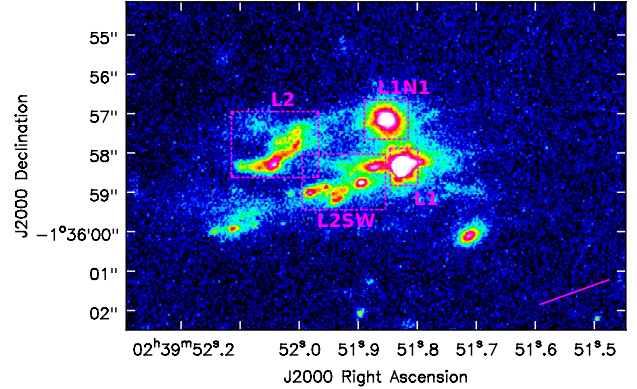}
\vspace{-4mm} 
\caption{The {\em HST} F814W (rest-frame 214 nm) image of SMM02399
  from the Frontier Fields archival data of the cluster Abell 370
  (Lotz et al. 2017).  The components are labeled with dotted-boxes as
  identified by Ivison et al. (2010).  The source is magnified by a
  factor of 2.38, and the line segment in the lower right indicates
  the lensing shear direction (Ivison et al. 2010).}
\end{figure}

\section{Observations}

\subsection{GBT Observations}

%
\begin{table}
  \caption{GBT Observations}
  \begin{tabular}{llccc}
    \hline \hline
   Date & GBT & Session &Zenith& On-Source\\ 
   (UT) &Project&       &Opacity&Time$^{a}$(s)\\
\hline
2016-04-05&  16A055 & 02&  0.032&  2815\\
2016-04-13&  16A055 & 03&  0.039&  4944\\
2016-04-23&  16A055 & 04&  0.057&  1824\\ 
2016-04-24&  16A055 & 05&  0.043&  9458\\
2017-10-07&  17B192 & 01&  0.071&  5056\\
2017-10-21&  17B192 & 02&  0.052&  6870\\
2017-10-27&  17B192 & 03&  0.044&  5313\\

\hline
 \end{tabular}

 \noindent $^{a}$The on-source time does not include the
 reference blank-sky scans or data flagged when the
 subreflector was in motion.
 
\end{table}

The GBT observations of SMM02399 were carried out in 2016 April and
2017 October.  In total, we collected 10.1 hours of on-source
integration time taken over seven observing sessions (Table~1).  The
observations targeted the \coa line at 30.2708\,GHz corresponding to a
redshift of $z=2.808$ (Frayer et al. 1998) using the Versatile GBT
Astronomical Spectrometer (VEGAS) with a bandwidth of 1080\,MHz and a
raw spectral resolution of 66\,kHz ($0.65\kps$).  We pointed the
telescope at the millimeter continuum centroid of (J2000) $02^{\rm
  h}39^{\rm m}51\fs87$, $-01\degr 35\arcmin 58\farcs8$ (Genzel et
al. 2003).

The observations were taken using sub-reflector beam switching
(``SubBeamNod'') with a 6 second switching period between the two
beams of the Ka-band receiver that are separated by 78$\arcsec$ on the
sky.  The beam size at the observed frequency is about 25$\arcsec$,
which is sufficient to cover all of the source components including
the large diffuse Ly$\alpha$ cloud.

The observations of broad, weak lines for single-dish telescopes are
limited by baseline stability, even for the GBT whose off-axis optics
avoids the strong standing waves from support structures that often
plague symmetric antennae.  The observing strategy was optimized to
minimize baseline issues and to produce an accurate \coa profile.  Two
spectral windows were offset by 150\,MHz to provide a check on the
baseline performance of the system downstream of the receiver.
Alternating SubBeamNod observations between the target and blank sky
were taken every 2 to 3 minutes to remove the residual baseline
structure.  We collected SubBeamNod scans of the bright nearby
pointing source (J0217+0144) before and after the pointing and focus
scans to track the loses due to pointing and focus drifts.

The absolute flux density scale was derived from observations of 3C48.
Based on the VLA calibration study of Perley \& Butler (2013), we
adopt a flux density of 0.91\,Jy for 3C48.  The uncertainty in
converting the observed GBT antenna temperature scale into flux
density is estimated to be 12\%.  This uncertainty includes the
uncertainty of the VLA calibration scale, measurement errors, the
uncertainty for the atmosphere correction, and the uncertainty
associated with the pointing and focus drifts.

\begin{figure}
\includegraphics[width=0.49\textwidth]{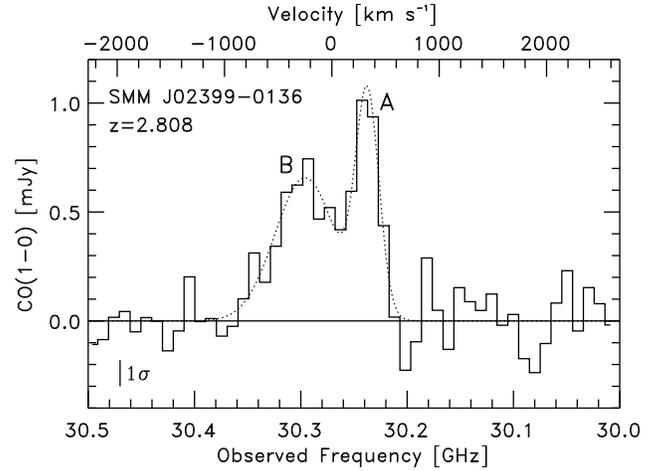}
\vspace{-4mm} 
\noindent{\small\addtolength{\baselineskip}{-3pt}} 
\caption{The GBT \coa spectrum of SMM02399.  The raw data have been
  smoothed to a channel resolution of $100\kps$.  The $1\sigma$ noise
  of 0.12\,mJy per channel is shown in the lower left.  The
  dotted-line shows the fit of a double Gaussian profile to the data.
  The velocity scale is based on a redshift of $z=2.808$.}

\end{figure}

\subsection{GBT Data Reduction} 

The GBT spectral-line data were reduced using GBTIDL (Marganian et
al. 2006).  After the standard reduction of the SubBeamNod scans,
there are significant residual baseline issues in the data on several
different frequency and time scales related to multiple instrumental
effects, including the different optical paths of the two subreflector
positions, the receiver itself, as well as issues with thermal
stability within the equipment room, which contains several key analog
components and VEGAS.

To mitigate the residual baseline structures, the blank-sky scans
taken immediately before and after each target scan were averaged and
used to derive a polynomial baseline that was removed from each target
scan before co-addition.  This method worked well for most of the
data, but scans still showing residual baseline structures with noise
levels higher than expected were deleted before co-addition.  The data
for each of the two beams and each of the two spectral windows were
processed independently before combining.  The resulting final
spectrum shows a broad-component of molecular gas (component B) and a
narrow component (component A) centered at $330\kps$ (Fig. 2).  The
GBT data were reduced using various methods to verify the results.
The shape of the \coa profile was confirmed using sub-sets of the data
for the different beams and different spectral windows.

\begin{table}
  \caption{ALMA Observations}
  \begin{tabular}{lclcc}
    \hline \hline
   Date &Number&Baselines & CO & On-Source\\ 
   (UT) &Antenna&[m]       &Transition&Time (s)\\
\hline
2014-06-13& 34 & 19--650& CO(7-6)&  816\\
2015-07-01& 34 & 43--1574&CO(3-2)& 1239\\
2015-09-05& 36 & 15--1574&CO(3-2)& 2328\\
2016-08-12& 36 & 15--1462&CO(3-2)& 2328\\
\hline

\end{tabular}

\end{table} 

\subsection{ALMA Observations}

The ALMA \coc and \cog observations of SMM02399 were carried out in
four observing sessions taken in Cycle-2 and Cycle-3.  The raw ALMA
visibilities were calibrated using CASA (McMullin et al. 2007) by the
North American ALMA Science Center.  The data were imaged using
natural-weighting to maximize the signal-to-noise (S/N) ratio.  The
spatial resolution of the \coc and \cog data are
$0\farcs7\times0\farcs6$ and $0\farcs6\times0\farcs5$, respectively.
The data were smoothed to produce image cubes with $100\kps$ and
$300\kps$ velocity resolution for analysis and comparison with the GBT
CO(1-0) data.  The noise level per $100\kps$ channel is 0.12\,mJy and
0.33\,mJy for the \coc and \cog data, respectively.

\begin{figure}
\includegraphics[width=0.49\textwidth]{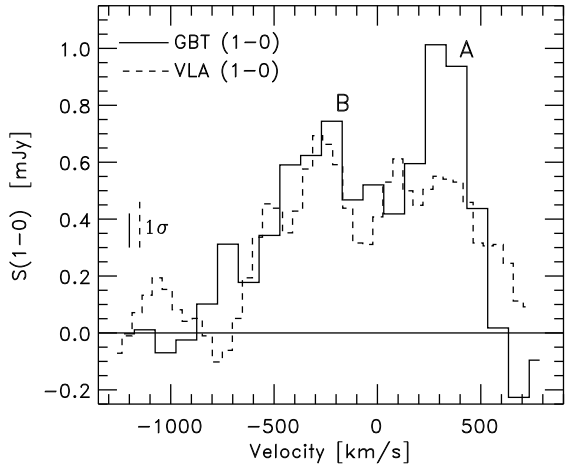}
\vspace{-4mm} 
\noindent{\small\addtolength{\baselineskip}{-3pt}} 
\caption{The GBT \coa spectrum compared with the published VLA \coa spectrum
  (Thomson et al. 2012).   The VLA spectrum has been scaled by 1.3 to
  match the total integrated line flux reported by the authors.   The
  GBT spectrum shows excess emission for component (A). 
  The $1\sigma$ noise levels are shown to the left.}
\end{figure}

\section{Results}

\subsection{GBT CO(1-0) Results}
\begin{table}
  \caption{\coa Measurements for Velocity Components}
  \begin{tabular}{ccccc}
    \hline \hline
 & Peak &    FWHM   &  Frequency & Velocity$^{a}$\\
 & [mJy]& [$\kps$]  & [MHz]      &  [$\kps$]\\
\hline

(A) & $1.00\pm 0.11$ &  $260\pm 35 $& $30237.4 \pm 1.4$ &  $330\pm 15$  \\
(B) & $0.66\pm 0.07$ &    $660\pm 100$ & $30295.9\pm 3.5$ &
$-250\pm 35$  \\

\hline
\end{tabular}
\noindent
Properties derived from fitting a double Gaussian to the \coa line
profile.  $^{a}$The velocity is with respect to a redshift of $z=2.808$.
\end{table}

The GBT \coa profile is consistent with two Gaussian components: (A) a
narrow component at $330 \kps$ and (B) a broad component centered at
$-250 \kps$ (Table~3).  Previously published CO profiles of SMM02399
(Frayer et al. 1998; Genzel et al. 2003; Thomson et al. 2012) also
showed a double peak profile.  However, for each of these
interferometric studies, the strength of the peak at negative
velocities was larger or similar to the strength of the peak observed
at positive velocities.  The GBT \coa profile shows a significant
excess for the positive velocity component (A) in comparison to the
VLA (Fig. 3), which highlights the importance of the single-dish
observations.  The narrow component (A) accounts for $38\pm7$\% of the
total \coa line flux.

Although the GBT profile shows excess emission for component (A), the
integrated \coa line flux from the GBT is similar to previous results
from the VLA within the uncertainties.  The integrated GBT \coa line
flux is $0.73\pm0.10 \jy \kps$.  The uncertainty includes the 12\%
calibration uncertainty combined in quadrature with the measurement
error.  In comparison with VLA observations, Thomson et al. (2012)
measured a \coa line flux of $0.60\pm0.12 \jy \kps$ and Ivison et
al. (2010) derived a value of $0.70\pm0.18 \jy \kps$.  Summing over
the channels in the VLA \coa profile shown in Thomson et al. (2012)
would suggest an integrated line flux of only $0.47 \jy \kps$ which is
less than their tabulated value of $0.60 \jy \kps$.  The higher value
is likely more appropriate based on the Ivison et al. (2010) results.
We scaled the VLA \coa spectrum from Thomson et al. (2012) to match
their tabulated line flux of $0.60 \jy \kps$ in Figure~3.  With this
scaling the the strength of velocity component (B) is consistent
between the GBT and the VLA.

Based on the \coa line flux and correcting for lensing, the intrinsic
\coa line luminosity for SMM02399 is
$L^{\prime}(\co)=(1.1\pm0.2)\times 10^{11} \Kkpspc$.  Depending on the
CO--to--H$_2$ conversion factor ($\alpha_{\rm CO}$), the derived CO
luminosity corresponds to a molecular gas mass of $(1.1\pm0.2) \times
10^{11} \alpha \msun$, where $\alpha$ is a unitless scalar
representing the $\alpha_{\rm CO}$ value.

\begin{figure}
\includegraphics[width=0.49\textwidth]{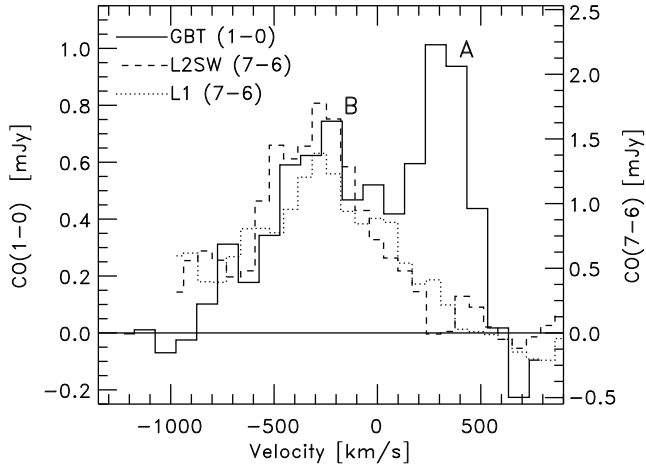}
\vspace{-4mm} 
\noindent{\small\addtolength{\baselineskip}{-3pt}} 
\caption{The ALMA \cog spectrum for L2SW and L1, shown as dashed and
  dotted lines respectively, compared to the GBT CO(1-0) spectrum.
  The \cog emission for both optical components L2SW and L1 is only
  associated with component (B) of the \coa profile and is not
  detected for component (A).  The \cog profiles at velocities less
  than $-700\kps$ are affected by [C{\sc i}] emission. }
\end{figure}

\subsection{ALMA CO Results}

The ALMA \cog observations revealed emission from the two central
optical components L2SW and L1.  The \cog velocity profiles from both
of these components are similar to velocity component (B) of the \coa
profile (Figure~4).  There is no evidence of any \cog emission
associated with velocity component (A).  This is an important result
that is fundamental to our understanding of the source.  Previously,
it has been speculated that the two peaks of the CO profile may arise
from the two optical components L1 and L2 of the merger system (e.g.,
Frayer et al. 1998) or represent a double-horn profile associated with
a molecular gas disk rotating around L1 (Genzel et al. 2003).  The
ALMA data rules out both of these interpretations.  Based on the \cog
data, the AGN (L1) and the red starburst region (L2SW) are at similar
systemic velocities corresponding to component (B) of the \coa
profile, and the narrow component (A) represents a distinct component
showing much lower CO excitation than that found from L1 and L2SW.  A
natural question is: where is the molecular gas associated with
velocity component (A) located?

\begin{figure}
\includegraphics[width=0.49\textwidth]{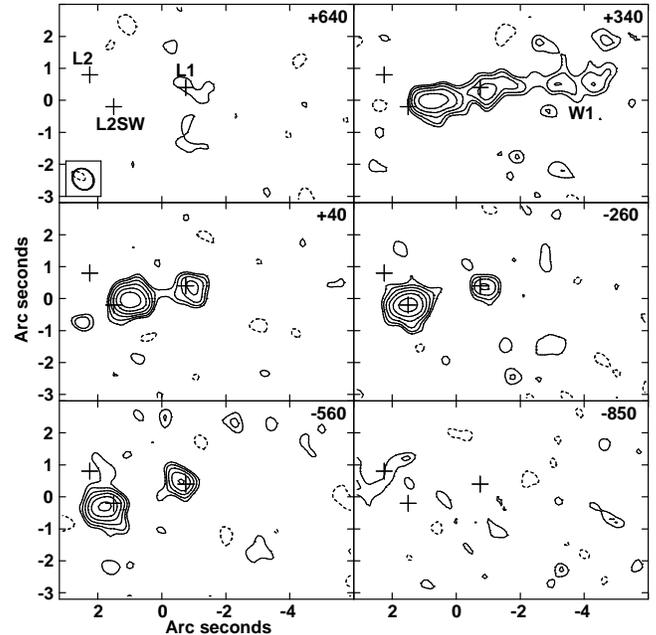}
\vspace{-4mm} 
\noindent{\small\addtolength{\baselineskip}{-3pt}} 
\caption{The ALMA CO(3-2) channel maps smoothed over $300\kps$.  The
  channel velocity in $\kps$ with respect to $z=2.808$ is shown in the
  upper-right of each panel.  The crosses mark the positions of the
  optical components L1, L2, and L2SW and are labeled in the first
  panel. The new extended western component is labeled as W1.  The
  $1\sigma$ noise level is $0.02\jy\kps$, and the contours start at
  $2\sigma$ and are incremented by $2^{0.5}$ ($-2$, 2, 2.83, 4, 5.66,
  8, 11.3$\sigma$).  The beam ($0\farcs72 \times 0\farcs61$) is shown
  in the lower left of the first panel.}
\end{figure}

Deep sub-arcsec CO(3-2) imaging with ALMA has helped to explain the
nature of component (A).  Figure~5 shows the ALMA CO(3-2) channel maps
smoothed over $300\kps$.  The three panels corresponding to center
velocities $+40$, $-260$, and $-560 \kps$ cover the velocity range
associated with component (B), while the panel centered on $+340\kps$
corresponds to component (A).  The $+340\kps$ panel shows a new
extended component of molecular gas (designated ``W1'') that extends
several arcsec westward from L2SW and L1.  The \coc spectrum for W1
has a velocity and line-width that closely matches component (A) of
the GBT CO(1-0) profile (Fig. 6).  The total linear extent of the
molecular gas for SMM02399 (L2SW+L1+W1) corresponds to $3\farcs2$
(25\,kpc) in the source-frame, after correcting for lensing which
shears the source emission roughly along the long axis of the observed
CO emission.

\begin{figure}
\includegraphics[width=0.49\textwidth]{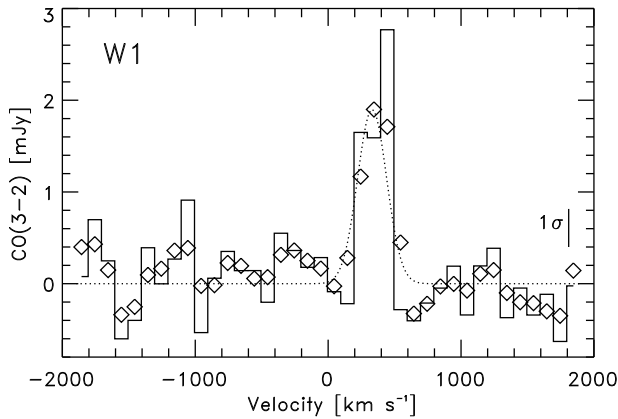}
\vspace{-4mm} 
\noindent{\small\addtolength{\baselineskip}{-3pt}} 
\caption{The ALMA CO(3-2) spectrum for the new W1 component of
  molecular gas.  The solid line show the spectrum with $100\kps$
  channels, and the $1\sigma$ error bar is given to the right. The
  Gaussian fit of component (A) of the GBT CO(1-0) profile (Fig. 2) is
  shown by the dotted line and has been normalized to the peak of the
  Hanning-smoothed CO(3-2) spectrum represented by the diamonds.}
\end{figure}

In addition to uncovering a new extended component of molecular gas
(W1), the ALMA CO(3-2) data show velocity gradients consistent with
rotation for both L2SW and L1.  By comparing panels $+40\kps$,
$-260\kps$, and $-560\kps$, an obvious velocity gradient across L2SW
is observed with a position angle of about $105\degr$ on the sky.  The
data also show evidence for a velocity gradient across L1.  A
straightforward interpretation of these observations is that both L2SW
and L1 are associated with separate disks of molecular gas undergoing
starburst activity.  These data will be analyzed and modeled in detail
in a future paper.  No molecular gas was detected from the L2 or L1N
optical components.

\begin{figure}
\includegraphics[width=0.49\textwidth]{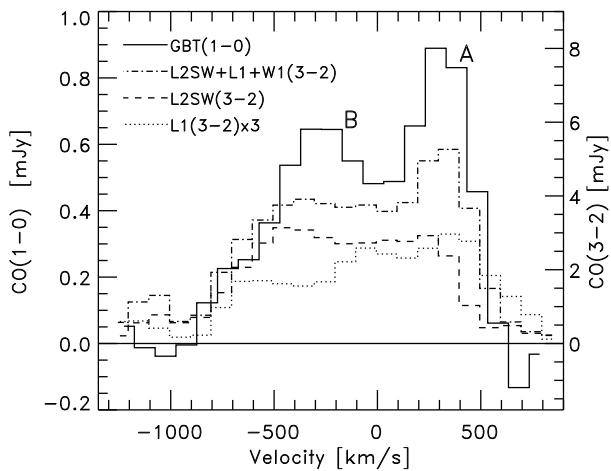}
\vspace{-4mm} 
\noindent{\small\addtolength{\baselineskip}{-3pt}} 
\caption{The total ALMA CO(3-2) spectrum, shown by the dashed-dotted
  line labeled as L2SW+L1+W1(3-2), compared to the GBT CO(1-0)
  spectrum (solid line).  The ALMA CO(3-2) spectra for L2SW and L1 are
  shown as dashed and dotted lines respectively.  All spectra have
  channels of $100\kps$ and have been Hanning-smoothed.  The spectrum
  for L1 has been scaled up by a factor of three for comparison
  purposes.  The \coc axis has been scaled to $r_{31}=1$ for
  comparison with \coa. }
\end{figure}

The CO(3-2) emission peaks on the starburst region of L2SW, and this
region is more than three times brighter in CO(3-2) than the AGN
component L1 (Fig. 7).  The CO(3-2) spectrum from only the central
L2SW and L1 regions does not show the excess CO emission seen at
positive velocities associated with component (A).  However, using an
aperture that also contains the new faint western feature W1, the
total CO(3-2) spectrum resembles the GBT \coa profile (Fig. 7).

\subsection{CO Excitation}

\begin{table}
\begin{center}
  \caption{CO Brightness Temperature Ratios}
  \begin{tabular}{ccc}
    \hline \hline
 & $r_{31}$ & $r_{71}$\\
\hline

(A) & $0.58\pm 0.11$ & $<0.009$ \\
(B) & $0.66\pm 0.13$ & $0.095\pm0.017$\\

\hline
\end{tabular}
\end{center}

\noindent The observed CO brightness temperature ratios for velocity components
(A) and (B), where $r_{ij} = T_{b}(i-[i-1])/T_{b}(j-[j-1])$.  The
upper-limit represents $3\sigma$, and the errors include the
flux calibration uncertainty.  The ratios are based on the
measured peaks of the two components.
\end{table}

The two velocity components of the GBT \coa profile show very
different CO excitation (Table~4), as highlighted by Figure~4.  With
the single-dish GBT \coa data, we cannot fully disentangle the CO
excitation spatially for the system, but we can measure the excitation
for the two velocity components.  Velocity component (B) is comprised
of L2SW and L1, while the emission from velocity component (A) comes
from L2SW, L1, and the new component W1.  To study the CO excitation
of velocity components (A) and (B), we used the radiative transfer
code RADEX (van der Tak et al. 2007).  We adopted typical
CO abundance ratios and velocity gradients used in
Large-Velocity-Gradient modeling for Galactic molecular gas (e.g., de
Jong et al. 1975) to compute the predicted CO spectral-line energy
distributions for a range of physical properties (Fig. 8).  Although
we have data for only three CO transitions, the observed line ratios
rule out significant parameter space.  The relatively high
CO(3-2)/CO(1-0) brightness temperature ratios of $r_{31} \ga 0.6$
imply densities larger than $10^{3} {\rm cm}^{-3}$, while the ratios
of $r_{71} < 0.1$ imply densities lower than $10^{5} {\rm cm}^{-3}$.

The modeling with RADEX and the observed $r_{71}$ ratio suggests that
the properties for component (B) are consistent with local starbursts
with densities of order $10^{4} {\rm cm}^{-3}$ (Kamenetzky et
al. 2014; Zhang et al. 2014; Mashian et al. 2016) and temperatures of
40--50\,K, which agrees with the inferred single-component dust
temperature of $41\pm1$\,K derived from {\em Herschel} photometry
measurements (Magnelli et al. 2012).  In contrast, component (A) shows
CO ratios consistent with lower temperatures and densities that are in
rough agreement with values found for molecular cloud complexes within
the Milky Way (Fixsen et al. 1999; Friesen et al. 2017).

The $r_{73}$ ratio associated with L1 is three times larger than that
observed for L2SW which implies different CO excitation in L1 and
L2SW.  Assuming the same $r_{31}$ ratio for L1 and L2SW as derived for
velocity component (B), then the implied \cog to \coa flux density
ratio for L1 is estimated to be $9\pm2$ (diamond symbol in Fig. 8).
This value is arguably a lower limit if $r_{31}$ is higher for L1 than
L2SW.  The inferred high $r_{71}$ ratio for L1 is consistent with high
excitation associated with X-ray emission from the AGN as seen in some
local ultraluminous infrared galaxies (ULIRGs) (van der Werf 2010;
Mashian et al. 2016).

\begin{figure}
\includegraphics[width=0.49\textwidth]{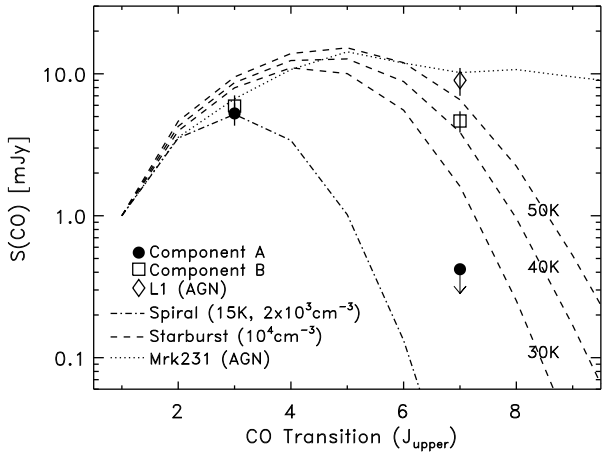}
\vspace{-4mm} 
\noindent{\small\addtolength{\baselineskip}{-3pt}}
\caption{The modeled CO spectral-line energy distribution normalized
  to CO(1-0) for a spiral galaxy with an excitation temperature of
  15\,K and density of $2\times 10^{3} {\rm cm}^{-3}$ shown as the
  dashed-dotted line while models for starbursts with a density of
  $10^{4} {\rm cm}^{-3}$ and temperatures of 30, 40, and 50\,K are
  shown as dashed lines.  The X-ray dominated AGN source Mrk\,231 (van
  der Werf et al. 2010) is shown by the dotted line.  The observed
  measurements for velocity components (A) and (B) are shown by the
  solid circles and open squares respectively.  Component (A) can be
  fitted by a spiral galaxy model while component (B) shows higher CO
  excitation consistent with a starburst.  The diamond symbol
  represents the estimated ratio for the AGN component L1.}
\end{figure}

\section{Discussion}

\subsection{Total Amount of Molecular Gas}

The single-dish \coa measurements from the GBT provide the best
constraints on the total amount of molecular gas in the system.  Since
the two velocity components show different CO excitation, they may
have different CO--to--H$_2$ conversion factors.\footnote{The
  molecular gas mass ($M[{\rm H}_{2}]$) derived by the CO--to--H$_2$
  conversion factor includes He.}  For the starburst component (B), we
adopt the standard ULIRG value of $\alpha_{\rm CO} =
0.8\msun(\Kkpspc)^{-1}$ (Downes \& Solomon 1998).  This value could
significantly underestimate the amount of gas arising from a high
density phase of the interstellar medium as discussed by Papadopoulos
et al. (2012), but the observed $r_{71}$ ratio for SMM02399 argues
against most of the molecular gas existing in very dense regions (\S
3.3).  The CO--to--H$_2$ conversion factor for component (A) is
arguably even more uncertain, but it is expected to be higher than the
value for the starburst component.  We adopt the Galactic value of
$4.3 \msun (\Kkpspc)^{-1}$ (Bolatto et al. 2013) for component (A) due
to its lower CO excitation.  If the molecular gas in component (A) has
low metalicity, the CO--to--H$_2$ conversion factor could be much
higher as found for metal-poor galaxies in the local universe (Rubio
et al. 1993; Wilson 1995; Israel 1997; Leroy et al. 2011; Schruba et
al. 2012; Amor\'{i}n et al. 2016).

Assuming the ULIRG CO--to--H$_2$ conversion factor for component (B)
and the Galactic conversion factor for component (A), SMM02399 has a
total molecular gas mass of $M({\rm H}_{2}) = (2.3\pm0.3)
\times10^{11} (\alpha/2.1) \msun$, with an effective single-value
CO--to--H$_2$ conversion factor of $\alpha_{\rm CO} =
2.1\msun(\Kkpspc)^{-1}$.  Velocity component (A) is responsible for about
40\% of the total \coa luminosity, but contains about 75\% of the
total molecular gas mass based on the adopted conversion factors.

\subsection{The Nature of W1}

The molecular gas mass associated with W1 that is spatially offset
from the central L2SW/L1 regions is estimated to be $M({\rm H}_{2}) =
(6.6\pm1.2)\times10^{10} (\alpha/4.3) \msun$, based on the fractional
CO(3-2) emission associated with this component, assuming the same
$r_{31}$ ratio for all gas in component (A), and using the adopted
CO--to--H$_2$ conversion factor.  This represents about 30\% of the
total molecular gas mass in the system.  A rough estimate for the
dynamical mass can be derived by assuming the gas in W1 is bound and
that the kinematics are dominated by isotropic turbulent motions.
With these assumptions, the dynamical mass is $M_{dyn} =2.3\times10^5
R_{\rm kpc} (3 \sigma_{v}^{2}) \msun$, where $\sigma_{v}$ is the
velocity dispersion in units of $\kps$.  W1 extends 13\,kpc from L1
($R_{\rm kpc} = 6.5$) and using the measured FWHM of component (A),
the dynamical mass of W1 is $M_{dyn} = (5.5\pm1.4)\times 10^{10}
\msun$.  These results are consistent with the mass of W1 being
dominated by molecular gas with a gas fraction near unity ($M({\rm
  H}_2)/M_{dyn} = 1.2\pm0.4 [\alpha/4.3] \approx 1$).  There is no
detected stellar component for W1 (Figure~1).  Aguirre et al. (2013)
derived the stellar masses of the optical components of SMM02399 based
on their colors and stellar population models.  Using these results
and assuming similar stellar populations for W1 as derived for the
nearby starburst region L2SW/L1sb, the $3\sigma$ upper limit on the
stellar mass for W1 is $M_{*} < 0.4 \times 10^{10} \msun$ based on the
noise level measured within the publicly available deep 1.6\,$\mu$m
Frontier Fields image (Lotz et al. 2017).  The large molecular gas to
stellar mass ratio of $M({\rm H}_2)/M_{*} > 16$ for W1 is more than a
factor of 100 larger than that found for normal local galaxies and 10
times larger than that derived for molecular-rich galaxies at
high-redshift (e.g., Popping et al. 2015, and references therein).

The exact nature of W1 is unclear.  We discuss three possible
explanations: (1) merger remnant, (2) outflow from the AGN, and (3)
inflowing gas.

The first explanation that W1 is a tidal remnant is arguably the most
natural explanation given the observed kinematics and source
morphology.  The kinematics of W1 appear to connect fairly smoothly
along the major axis of the massive molecular disk associated with
L2SW.  However, most of the gas associated with mergers is expected to
be concentrated in the central regions and not in tidal debris.  The
numerical simulations of Barnes (2002) show that the central regions
of gas-rich mergers are expected to contain 85-95\% of the gas mass,
while the surrounding loops and tidal tails are expected to contain
only 5-15\% of the gas mass.  The derived 30\% gas mass fraction for
W1 is higher than these predictions, but SMM02399 could still be in
the early stages of the merger event whereby much of the mass of W1
may fall back into the central regions.  Also, the details of the
merger orientation and kinematics are important.  If L2SW and L1 are
undergoing a prograde interaction within the same plane, we could
expect a tidal arm feature at the position of W1 that is amplified
based on the modeling of Oh et al. (2015), which may explain the
higher than expected mass fraction for W1.

A second possible explanation is that W1 is due to a massive molecular
outflow associated with the powerful AGN.  SMM02399 has a
significantly lower infrared-to-radio luminosity ratio than that found
for starbursts implying the importance of the AGN for the radio
emission (Frayer et al. 1998).  Also, the radio emission extends over
7$\arcsec$ in the east-west direction between L1 and L2/L2SW (Ivison
et al. 2010) and is aligned with the direction of the W1 emission.
The excess radio emission and its alignment along the direction of the
W1 emission may suggest a physical connection between the AGN and the
W1 molecular gas.  The observed GBT \coa profile also appears very
similar to predicted single-dish CO profiles from gas-rich merger
models that include AGN feedback with outflows (Narayanan et
al. 2006).  These simulations predict a narrow velocity-width CO
component due to the outflow that is offset in velocity from the
underlying broad velocity component arising from the central
star-forming regions.  The narrow CO line widths of these modeled
outflows appear consistent with slow moving molecular outflows as seen
in local starbursts, such as M82 (Walter et al. 2002), instead of the
fast moving outflows seen locally for powerful AGN that show very
broad CO velocity profiles (e.g., Mrk~231, Feruglio et al. 2010).

In the merger simulations of Narayanan et al. (2006, 2008), the
typical outflow mass is predicted to represent 5--15\% of the total
molecular gas mass, which is less than the 30\% mass fraction
estimated for W1.  However, other models have predicted higher mass
fractions for AGN dominated outflows in high-redshift galaxies in
rough agreement with W1 (e.g., Biernacki \& Teyssier 2018).  If W1 is
an outflow, the power associated with the kinetic energy of the
outflow is about $9\times 10^{10}\lsun$ which is about 1\% of the
bolometric luminosity of the system and well within the realm of
possibility.

A third explanation for W1 is that this molecular gas is inflowing
material associated with the formation of the galaxy.  Cosmological
models for the formation of SMGs predict massive gaseous inflows and
clumps of gas infalling into the central regions that build up the
mass of the galaxy over time (Dav\'{e} et al. 2010; Narayanan et
al. 2015).  The early inflowing material is expected to be metal-poor,
and if so, then the mass of molecular gas could be 10 times larger
than that estimated in \S~4.1 due to a higher CO--to--H$_2$ conversion
factor.  For metal-poor gas, CO is not a good tracer of the total
molecular gas mass due to the photodissociation of CO molecules from a
lack of shielding (Wolfire et al. 2010).  If the bulk of the molecular
gas in SMM02399 is metal-poor infalling material, then we would expect
to find strong atomic [C\,{\sc i}] emission in comparison to CO.  The
lack of a strong enhancement of [C\,{\sc i}] emission for SMM02399
(Walter et al. 2011) argues strongly against the scenario of W1
resulting from metal-poor inflowing gas.  However, W1 may be inflowing
metal-enriched gas.  For example, recent observations have suggested
the existence of metal-enriched gas within the molecular
circumgalactic medium of the $z=2.2$ Spiderweb (Emonts et al. 2018).

\section{Concluding Remarks}

In this paper, we present GBT and ALMA CO observations of the $z=2.8$
SMG SMM02399.  The GBT observations uncovered a component of excess
\coa emission.  This component has the same velocity and line-width as
a new faint structure (W1) imaged in CO(3-2) extending 13\,kpc
westward from the AGN.  The exact nature of W1 is currently unclear,
but it is estimated to contain about 30\% of the total molecular gas
in the system.

The results for SMM02399 highlight the potential contribution that
single-dish observations can have for studying molecular emission on
large spatial scales for high-redshift galaxies.  In general, most
high-redshift CO interferometric observations are limited to detecting
only the bright central cores of young galaxies.  The amount of
molecular gas on large spatial scales is currently unclear, although
models predict a significant amount of gas on the scale of galaxy
halos ($\sim 100$\,kpc) for galaxies in their formative phases
(Narayanan et al. 2015), and recent observations support these models
by detecting molecular gas on large scales in the Spiderweb system
(Emonts et al. 2016; 2018).  Single-dish observations of high-redshift
\coa, in particular with the GBT (Hainline et al. 2006; Swinbank et
al. 2010; Frayer et al. 2011; Harris et al. 2012; Harrington et
al. 2018), can test for the presence of excess CO emission and measure
the total molecular gas reservoirs of young systems.

\section*{Acknowledgments}

We thank our colleagues at the Green Bank Observatory who have made
these observations possible.  We greatly appreciate the staff at the
North American ALMA Science Center who calibrated the raw ALMA
visibilities.  The Green Bank Observatory and the National Radio Astronomy
Observatory are facilities of the National Science Foundation operated
under cooperative agreements by Associated Universities, Inc. 

This paper makes use of the following ALMA data: ADS/JAO.ALMA
\#2013.1.00403.S, ADS/JAO.ALMA \#2015.1.00302.S.  ALMA is a
partnership of ESO (representing its member states), NSF (USA) and
NINS (Japan), together with NRC (Canada), MOST and ASIAA (Taiwan), and
KASI (Republic of Korea), in cooperation with the Republic of Chile.
The Joint ALMA Observatory is operated by ESO, AUI/NRAO, and NAOJ.
The optical image is based on observations obtained with the NASA/ESA
Hubble Space Telescope, retrieved from the Mikulski Archive for Space
Telescopes (MAST) at the Space Telescope Science Institute (STScI).
STScI is operated by the Association of Universities for Research in
Astronomy, Inc. under NASA contract NAS 5-26555.  RJI acknowledges ERC
funding in the form of Advance Grant COSMICISM (321302). IRS
acknowledges support from the ERC Advance Grant DUSTYGAL (321334), a
Royal Society/Wolfson Award, and STFC(ST/P000541/1).

\facilities{GBT, ALMA}

\software{RADEX (van der Tak et al. 2007), GBTIDL (Marganian et
  al. 2006), CASA (McMullin et al. 2007)}



\end{document}